\documentclass[a4paper]{extarticle}
\usepackage[T1]{fontenc} 
\usepackage[english]{babel} 
\usepackage{indentfirst}

\usepackage[hmarginratio=1:1,left=20mm,right=20mm,bottom=25mm,top=25mm,columnsep=20pt]{geometry}

\usepackage[hang,normalsize,labelfont=bf,up,textfont=it,up,skip=0pt,justification=justified]{caption} 

\usepackage{abstract}

\usepackage{titlesec}
\titleformat{\section}[block]{\Large\bfseries\centering}{\thesection.}{1em}{}
\titleformat{\subsection}[block]{\large\bfseries}{\thesubsection.}{1em}{}
\titleformat{\subsubsection}[block]{\normalsize\bfseries}{\thesubsubsection.}{1em}{}

\usepackage{fancyhdr}
\renewcommand{\headrulewidth}{0pt}
\renewcommand{\footrulewidth}{0pt}

\usepackage{titling} 

\usepackage[breaklinks=true]{hyperref} 

\usepackage{amsmath,amssymb}
\usepackage{graphicx}
\usepackage{multirow}
\usepackage{algorithm}
\usepackage{stmaryrd}
\usepackage{empheq}
\usepackage[noend]{algpseudocode}
\usepackage{color}
\usepackage{bm}
\usepackage{lastpage}
\usepackage{longtable}

\makeatletter

\newcounter{propcounter}

\newcounter{remcounter}

\newcommand{\qed}{\nobreak \ifvmode \relax \else
      \ifdim\lastskip<1.5em \hskip-\lastskip
      \hskip1.5em plus0em minus0.5em \fi \nobreak
      \vrule height0.75em width0.5em depth0.25em\fi}

\renewcommand{\qed}{\nobreak \ifvmode \relax \else
      \ifdim\lastskip<1.5em \hskip-\lastskip
      \hskip1.5em plus0em minus0.5em \fi \nobreak
      \vrule height0.75em width0.5em depth0.25em\fi}

\newcommand{\ArtTitle}[2]{\def\@ArtTitle{#1}\def\@ArtTitleShort{#2}}
\newcommand{\ArtJournalInfo}[1]{\def\@ArtJournalInfo{#1}}
\newcommand{\ArtJournalRef}[1]{\def\@ArtJournalRef{#1}}
\newcommand{\ArtAuthors}[9]{
\def\@ArtAuthorA{#1}\def\@ArtAuthorAmail{#2}\def\@ArtAuthorAAff{#3}
\def\@ArtAuthorB{#4}\def\@ArtAuthorBmail{#5}\def\@ArtAuthorBAff{#6}
\def\@ArtAuthorC{#7}\def\@ArtAuthorCmail{#8}\def\@ArtAuthorCAff{#9}
}
\newcommand{\ArtAuthorsBis}[9]{
\def\@ArtAuthorD{#1}\def\@ArtAuthorDmail{#2}\def\@ArtAuthorDAff{#3}
\def\@ArtAuthorE{#4}\def\@ArtAuthorEmail{#5}\def\@ArtAuthorEAff{#6}
\def\@ArtAuthorF{#7}\def\@ArtAuthorFmail{#8}\def\@ArtAuthorFAff{#9}
}
\newcommand{\ArtAbstract}[1]{\def\@ArtAbstract{#1}}

\newcommand{\makeArtTitle}{

{
\includegraphics[height=0.1\textheight,trim={4cm 12cm 4cm 12cm},clip]{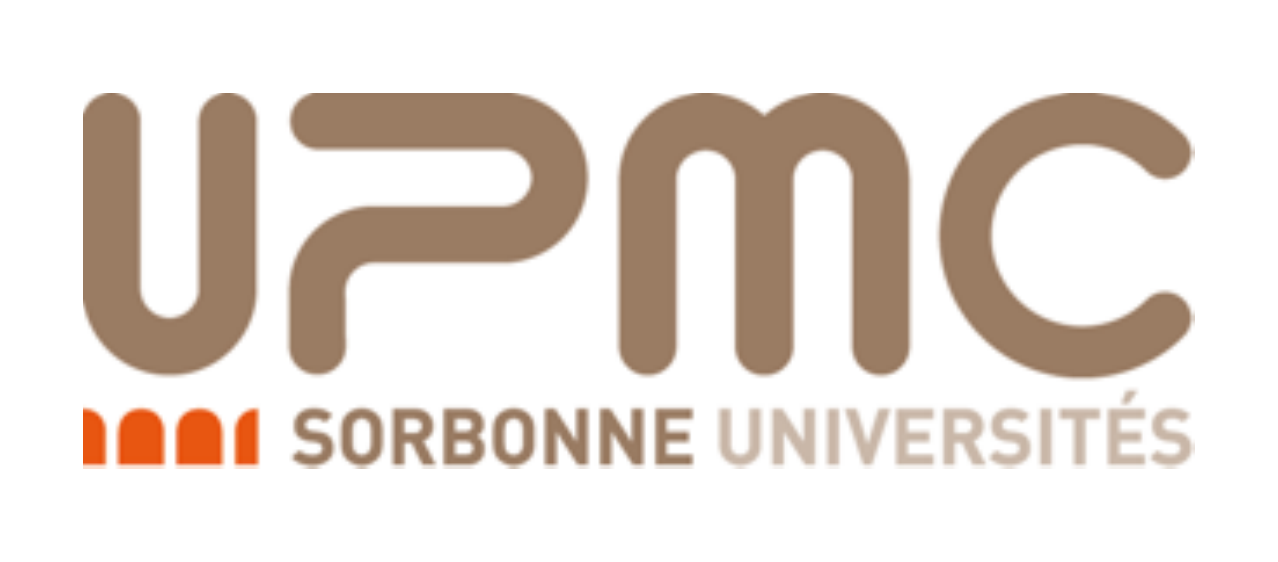}
\hspace{0.15\textwidth}
\includegraphics[height=0.1\textheight,trim={9cm 13cm 9cm 13cm},clip]{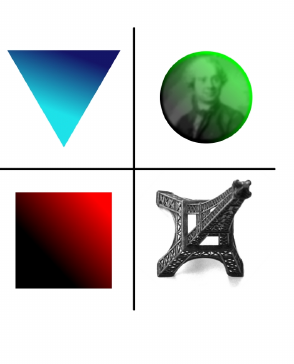}
\hspace{0.15\textwidth}
\includegraphics[height=0.1\textheight,trim={6cm 10cm 6cm 10cm},clip]{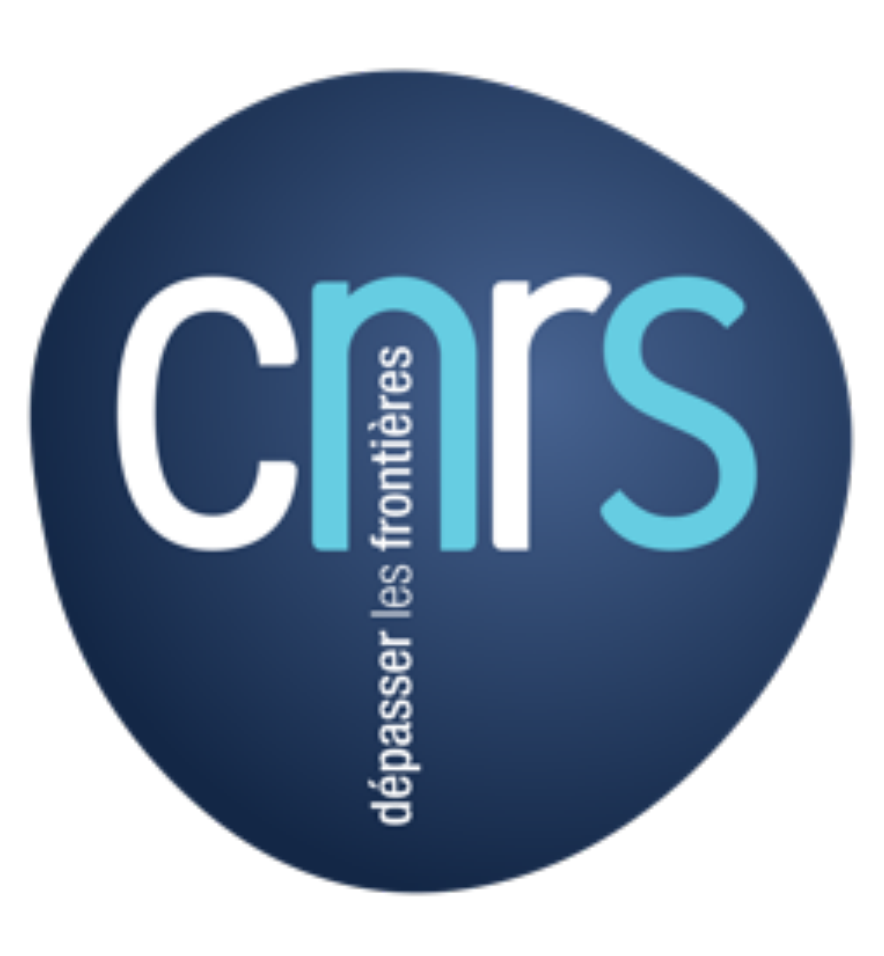}
}

\par\vspace{0.05\textheight}

{
\begin{minipage}{0.33\textwidth}
\centering
\Large\@ArtAuthorA
\end{minipage}
\begin{minipage}{0.33\textwidth}
\centering
\Large\@ArtAuthorD
\end{minipage}
\begin{minipage}{0.33\textwidth}
\centering
\Large\@ArtAuthorE
\end{minipage}
\hfill
\par\vspace{0.01\textheight}
\begin{minipage}{0.49\textwidth}
\centering
\Large\@ArtAuthorC
\end{minipage}
\begin{minipage}{0.49\textwidth}
\centering
\Large\@ArtAuthorB
\end{minipage}
}
\par\vspace{0.0\textheight}
{\centering 
\small\@ArtAuthorCAff\\
}
\par\vspace{0.05\textheight}

\begin{center}
\begin{minipage}{0.9\textwidth}
\centering
\huge
\@ArtTitle
\end{minipage}
\end{center}
\par\vspace{0.05\textheight}

{
{\Large\centering\textbf{Abstract\\}}
\par\vspace{0.02\textheight}
{\noindent
\textit{\@ArtAbstract}
}
}
\par

\par\vspace*{\fill}
{\huge
\href{http://www.arXiv.org}{\nolinkurl arXiv} - \href{http://hal.upmc.fr}{\nolinkurl HAL}
}

\renewcommand{\footrulewidth}{1pt}
\fancyfoot{
\raggedright
\@ArtJournalInfo
\@ArtJournalRef
\\
\underline{Contact}: \@ArtAuthorAmail
}

\pagebreak
}

\newcommand{\makeArtTableOfContent}{
\tableofcontents
\pagebreak
}

\makeatother

\ArtTitle{
One-dimensional Arterial Network Model for Bypass Grafts Assessment
}{
1D Bypass Grafts assessment
}
\ArtJournalInfo{
In revision in \textit{Medical Engineering \& Physics}
}
\ArtJournalRef{\url{}}

\ArtAuthors
{Arthur R. Ghigo}{arthur.ghigo@dalembert.upmc.fr}
{Sorbonne Universit\'es, CNRS and UPMC Universit\'e Paris 06, UMR 7190, Institut Jean Le Rond $\partial$'Alembert}
{Jose-Maria Fullana}{}
{Sorbonne Universit\'es, CNRS and UPMC Universit\'e Paris 06, UMR 7190, Institut Jean Le Rond $\partial$'Alembert}
{Pierre-Yves Lagr\'ee}{}
{Sorbonne Universit\'es, CNRS and UPMC Universit\'e Paris 06, UMR 7190, Institut Jean Le Rond $\partial$'Alembert}

\ArtAuthorsBis
{Salam Abou Taam}{}
{Sorbonne Universit\'es, CNRS and UPMC Universit\'e Paris 06, UMR 7190, Institut Jean Le Rond $\partial$'Alembert}
{Xiaofei Wang}{}
{Sorbonne Universit\'es, CNRS and UPMC Universit\'e Paris 06, UMR 7190, Institut Jean Le Rond $\partial$'Alembert}
{Pierre-Yves Lagr\'ee}{}
{Sorbonne Universit\'es, CNRS and UPMC Universit\'e Paris 06, UMR 7190, Institut Jean Le Rond $\partial$'Alembert}

\ArtAbstract
{
We propose an arterial network model based on 1D blood hemodynamic equations to study the behavior of different vascular surgical bypass grafts in case of an arterial occlusive pathology: an obliteration or stenosis of the iliac artery. We investigate the performances of three different bypass grafts (Aorto-Femoral, Axillo-Femoral and cross-over Femoral) depending on the degree of obliteration of the stenosis. Numerical simulations show that all bypass grafts are efficient since we retrieve in all cases the normal hemodynamics in the stenosed region while ensuring at the same time a global healthy circulation. We analyze in particular the Axillo-Femoral bypass graft by performing hundreds of simulations by varying the values of the Young's modulus [$0.1 - 50 \ MPa$] and the radius [$0.01 - 5 \ cm$] of the bypass graft. We show that the Young's modulus and radius of commercial bypass grafts are optimal in terms of hemodynamic considerations. The numerical findings prove that this approach could be used to optimize or plan surgeries for specific patients, to perform parameter's statistics and error propagation evaluations by running extensive simulations, or to characterize numerically specific bypass grafts. 
}

\newcommand{\myreferences}{mybibfile.bib}

\begin{document}


\makeArtTitle


\makeatletter
\fancypagestyle{plain}{
\fancyhf{}

\renewcommand{\headrulewidth}{1pt}
\fancyhead[L]{\@ArtTitleShort \: - \: \@ArtAuthorA \, et al.}
\fancyhead[R]{\thepage /\pageref{LastPage}} 

\renewcommand{\footrulewidth}{0pt}
\fancyfoot{}
}
\makeatother


\makeArtTableOfContent


\makeatletter
\fancyhf{}

\renewcommand{\headrulewidth}{1pt}
\fancyhead[L]{\@ArtTitleShort \: - \: \@ArtAuthorA \, et al.}
\fancyhead[R]{\thepage /\pageref{LastPage}} 

\renewcommand{\footrulewidth}{0pt}
\fancyfoot{}

\makeatother


\section{Introduction}

Arterial diseases are frequent clinical pathologies, and their prevalence is evaluated from 3$\%$ to 10$\%$ in the global population with a significant growth in persons over 70 years old (from 15$\%$ to 20$\%$) \cite{Norgren2007}. Some of these pathologies cause symptoms going from intermittent claudication to severe ischemia. These symptoms are due to a decrease in blood supply when the vessels providing vascularization are narrowed or occluded. These arterial lesions can have severe consequences when they occur in the lower members, such as in the Iliac arteries, where they can lead to the amputation of the stenosed member if the blood flow is not restored in time.

When the symptoms are too severe or when medical treatment fails, surgery is necessary to restore the flow of blood in the stenosed member. This can be done by angioplasty stenting, where the obliterated segment is replaced by a prosthesis (stent) during an endovascular substitution surgery. An alternative solution consists in inserting an anastomosis graft to redirect the flow of blood from a healthy artery to bypass the obliterated vessel and restore the blood flow after the stenosis. In both cases, the mechanical role of these grafts or conduits is to  replace or bypass vessels that  have  become  occluded or severely obstructed by a disease processes \cite{Abbott1993}.
  
Numerical studies of local endovascular graft replacements have been done in \cite{Marchandise2009,Willemet2013}.
We propose to study instead extracorporeal bypass graft procedures using a detailed model of the systemic network. We compute the blood flow in each segment of the network to evaluate the performances of extracorporeal bypass grafts in case of occlusive pathologies. The particular pathology we study is a stenosis of the Right Iliac artery, which can be defined as a severe obliteration or narrowing of the cross-sectional area of the artery, resulting in a small or negligible flow rate after the obliteration. 

In the pathological case of an Iliac artery obliteration, the most common bypass graft configurations are: Aorto-Femoral, Axillo-Femoral and cross-over Femoral, defined by the names of the healthy or donor artery (Aorto for Aorta, Axillo for Axillary and cross-over standing for the same artery on the opposite leg) and the name of the receptor artery, in our case the Right Femoral artery which follows distally the narrowed site. In clinical routines, the Aorto-Femoral bypass graft is often preferred. However, for weak patients who can not tolerate the aortic clamping required to insert the Aorto-Femoral bypass graft, the preferred solution is the extra-anatomic Axillo-Femoral bypass graft \cite{Appleton2010}.

The aim of this communication is to present a numerical model of an arterial network presenting a stenosis of the Right Iliac artery and to compute numerical solutions of the blood flow after the bypass graft surgery is performed. To help the medical staff optimize surgical repair, we evaluate the flow rate and pressure after the bypass graft and in the opposite member (Left Iliac artery), which is an a posteriori evaluation  of the quality of the surgery. We study in particular the optimization of the Axillo-Femoral bypass graft because it is proposed for patients with fragile health who could not withstand an additional surgical intervention and because it has the less graft survival time among the three previously named bypass grafts (\cite{Greenwald2000,Musicant2003}). We hope that this numerical approach will also be used to optimize new prosthesis.
It is important to keep in mind that the prevalence of the lower extremity arterial diseases increases with age, from 3\% for those under 60 years old to over 20\% for those over 75 years old (\cite{Vogt1992,Weitz1996}). Therefore, a numerical tool for planning surgeries could be an interesting protocol step in the future. 

In the following, we propose only hemodynamic predictions based on fluid mechanics equations, regardless of the biological phenomena and their consequences. Nevertheless, we are aware that short term graft failures can be caused by infections or hemorrhages, while long-term failures  are the result of intimal hyperplasia of the graft site, with a proliferation and a migration of vascular smooth muscle cells near the arterial wall \cite{Greenwald2000}.

In Section 2, we present the numerical model and the model arterial network. The pathological network, the numerical results for the three bypass grafts and the parametric study of the Axillo-Femoral bypass graft are presented in Section 3, where we also discuss the numerical findings.

\section{Numerical model}

To compute the blood dynamics in an artery, we use a set of 1D equations expressed in terms of the dynamical variables of flow rate $Q$, cross-sectional area $A$ and internal average pressure $P$. This 1D system of equations results from the integration over the cross-sectional area of the Navier-Stokes equations for an incompressible Newtonian fluid, giving the following mass and momentum 1D conservation equations
 \begin{eqnarray}
\frac{\partial A}{\partial t}+\frac{\partial Q}{\partial
x}=0,
\label{massConserv_AQ}\\
\frac{\partial Q}{\partial t}+ \frac{\partial}{\partial x}(
\frac{Q^2}{A})+
\frac{A}{\rho}\frac{\partial P}{\partial x}=
2\pi \nu \left [ r \frac{\partial v_x}{\partial r} \right
]_{r=R}, 
\label{momentumConserv_AQ}
\end{eqnarray} 
where $v_x$ is the axial velocity, $\rho$ is the fluid density and $\nu$ is the kinematic viscosity of the fluid. We use the typical blood values $\rho = 1 \ g / cm^3$ and $\nu = 3.5 \ 10^{-2} \ cm^2 / s$. The internal pressure $P$ is related to the cross-sectional area $A$ by 
\begin{eqnarray} 
P=P_{ext}+\beta(\sqrt{A}-\sqrt{A_0})+\nu_s \frac{\partial
A}{\partial t},\label{constitutive}
\end{eqnarray}
assuming that the arterial wall is thin, isotropic, homogeneous, incompressible and that it deforms axisymmetrically with each circular cross-section independently of the others. $\beta$ and $\nu_s$ are the wall's coefficients for elasticity and viscoelasticity respectively, and we use a Kelvin-Voigt model to describe the viscoelastic behavior of the wall \cite{Alastruey2011}. 

By approximating the friction drag by $-C_f Q/A$ and using the expression \eqref{constitutive} for the pressure $P$, we can re-write the momentum equation  as 
\begin{eqnarray}
\frac{\partial Q}{\partial t}+
\frac{\partial}{\partial x}\bigl(\frac{Q^2}{A}+\frac{\beta}{3\rho}A^\frac{3}{2}\bigr)=
-C_f\frac{Q}{A}+C_v\frac {\partial ^2 Q}{\partial^2 x}. 
\label{momentumConserv_AQ2}
\end{eqnarray} 
 We set $C_f=22\pi\nu$ as was computed from coronary blood flow in Smith et al.~\cite{Smith2002}.
It is straightforward to see that the coefficients 
$$\beta=\frac{\sqrt{\pi} Eh}{(1-\eta^2)A_0}$$ 
and 
 $$C_v=\frac{\sqrt{\pi}\phi h}{2\rho(1-\eta^2)\sqrt{A_0}}$$ 
 in equation (\ref{momentumConserv_AQ2}) respectively incorporate the elastic and the viscoelastic behaviors of the wall dynamics of the coupled fluid-structure interaction problem. In this approach, we define $C_v=\frac{A\nu_s }{\rho}$ and we use the following wall parameters, presented in Table \ref{tab:systemic_network} in  \ref{annexe} for the systemic network model we consider: the Young’s modulus $E$, the Poisson ratio $\eta$, the viscoelastic coefficient $\phi$ and the arterial thickness $h$.  More details can be found in reference \cite{Wang2016}.

From a mathematical point of view, the system of equations here presented is composed of a hyperbolic (transport equation) and a parabolic part (viscoelastic source term). To obtain the numerical solution of both parts, we introduce a mesh in the axial direction by dividing each artery in a series of cells of size $\Delta x$. When then define the discrete time $t^n = n \Delta t $, where $\Delta t$ is the time step. Using this decomposition of the space and time domains, we discretize the hyperbolic part with a MUSCL (monotonic upwind scheme for conservation law) finite volume scheme and the parabolic part with a Crank-Nicolson scheme. We compute the numerical solution using a code developed by our laboratory and written in C++ and parallelized with OpenMP. The numerical implementation of the full viscoelastic nonlinear system has been validated by comparing the computed solutions to analytic solutions of the linearized system and experimental data~\cite{Saito2011,Wang2016}.

The network used in the numerical simulations is constructed by connecting different arterial segments together. The connections take place at branching points, where mass conservation and dynamical pressure continuity are imposed. As an explanatory example we considered a simple problem: a parent vessel connected to two daughter
arteries. At the branching point, there are then six unknowns at the iteration $n+1$: $A_p^{n+1}$ and $Q_p^{n+1}$ for the outlet of the parent
artery and $A_{d_1}^{n+1}$, $Q_{d_1}^{n+1}$,$A_{d_2}^{n+1}$ and
$Q_{d_2}^{n+1}$ for the inlets of the two daughter arteries (numerically speaking, $n$ refers to time $t^n$ and $n+1$ to time $t^{n+1}$). 
All these quantities are function of the  values at the current iteration $n$. The pressures for the parent and daughter arteries, respectively $P_p^{n+1}$ and $P_{d_i}^{n+1}$, are expressed as a function of the cross-sectional area $A$ using the constitutive relation linking the pressure and the cross-sectional area (equation (\ref{constitutive})). From the physical
point  of view, we preserve the basic laws of conservation, that is the conservation of mass flux
\begin{eqnarray*}
Q_p^{n+1}-Q_{d_1}^{n+1}-Q_{d_2}^{n+1}=0,     \label{equ1}
\end{eqnarray*}
and the conservation of momentum flux
\begin{eqnarray*}
\frac{1}{2}\rho \left(\frac{Q_p^{n+1}}{A_p^{n+1}}\right)^2+P_p^{n+1}-\frac{1}{2}\rho \left(\frac{Q_{d_i}^{n+1}}{A_{d_i}^{n+1}}\right)^2-P_{d_i}^{n+1}=0  \label{equ2}
\end{eqnarray*}
 There should be some additional terms to take into account energy losses due to the complex flow in the connection sites but, in practice, these losses only have secondary effects on the pulse waves \cite{Alastruey2011}, and we neglect them. The last three equations we need to complete the resolution of the branching point problem come from the matching at the conjunction point of incoming and outgoing characteristics of the hyperbolic problem.
 
The boundary conditions for the input and the output sites of the network are: (i) input: an imposed physiological flow rate at the principal artery starting from the heart and (ii) output: reflection coefficients characterizing the resistance of the vascular bed that is not taken into account in our model. Theses values are shown in the last column of Table \ref{tab:systemic_network} in  \ref{annexe}. To define the input boundary condition in terms of the blood flow rate, we introduce the ejection fraction $EF$, which is essentially the ratio of the difference between the End Diastolic Volume (EDV) and the End Systolic Volume ESV over the EDV:
\begin{eqnarray}
EF = { EDV - ESV \over EDV} \times 100 .
\end{eqnarray}
Healthy people typically have ejection fractions between $50\%$ and $65\%$. On the contrary, people with heart muscles damages (principally on the myocardium) have a low ejection fraction. The input signal used in the numerical simulations is the following
\[  Q(t)= \left\{ \begin{array}{ll}
         Q_{max} \sin ( { 2 \pi \over T} t) & \mbox{if $t \leq T/2$};\\
        0. & \mbox{if $t > T/2$}.\end{array} \right. \] 
Therefore, the ejected volume $V_e=EDV-ESV$ per period $T$ is computed by integrating $Q(t)$ over a period, 
\begin{eqnarray}
V_e = Q_{max} { T \over \pi}.
\end{eqnarray}
We find then that we can impose $Q_{max}$ as 
\begin{eqnarray}
Q_{max}  = EF \ \pi \ {EDV \over T}
\label{eq:Q}
\end{eqnarray}
and we can now define, for a given period $T$ and a given End Systolic Volume ($EDV$), $Q_{max}$ as a function of the ejection fraction $EF$. With this approach we constructed a simple heart model that allows us to define a pathological heart by reducing the ejection fraction. We note that the last relation has a physical meaning as long as in the case of a malfunctioning heart, the system reacts by either increasing the $EDV$ by expanding the muscular fibers or by reducing the period $T$ by increasing the cardiac rhythm. Both situations are modeled by equation (\ref{eq:Q}).

\section{Methods and results}

In this section we present the numerical protocol followed as well as the numerical results obtained with the three different bypass grafts (Axillo-femoral, Femoral-Femoral and cross-over Femoral bypass grafts) applied to a pathological network presenting an obliteration of the Right Iliac artery. The numerical protocol is the following:
\begin{enumerate}
	\item we first simulate  a ``healthy'' network (Figure \ref{fig:Artree} (a)). The computed numerical data will be used as the target that the pathological network treated with extracorporeal bypass grafts should aim for,
	\item  we then build the pathological network by narrowing the principal artery of the lower leg, the Right Iliac artery (Figure \ref{fig:Artree} (b)). As we have access to all the hemodynamic variables everywhere in the network we will be able to observe the global changes in terms of blood pressure and blood flow rate, depending on the degree of obliteration of the stenosis. We now have data for the pathological state for different degrees of obliteration,
	\item finally, we build the ``repaired'' network by modeling the Axillo-Femoral, Femoral-Femoral and cross-over Femoral bypass grafts using an elastic tube inserted between the donor and the receptor arteries of each bypass graft in the pathological network (Figure \ref{fig:Artree} (c) for the Axillo-Femoral, the other two are presented in Figure \ref{fig:bypasses}). The numerical results of the ``repaired'' network are then compared to the pathological and healthy data. 
\end{enumerate}

The key points of clinical repair are first the ability of the bypass graft to restore the blood flow in the previously non-perfused region (here the Right Femoral artery), and second, ensuring that the repair does not ill-balance the rest of the hemodynamic circulation. In the following, both key points are checked systematically for each bypass graft considered.

\subsection{Healthy state}

\begin{figure}[htb]
	\centering
	\includegraphics[width=0.7\textwidth, trim= 1cm 1cm 1cm 1cm, clip]{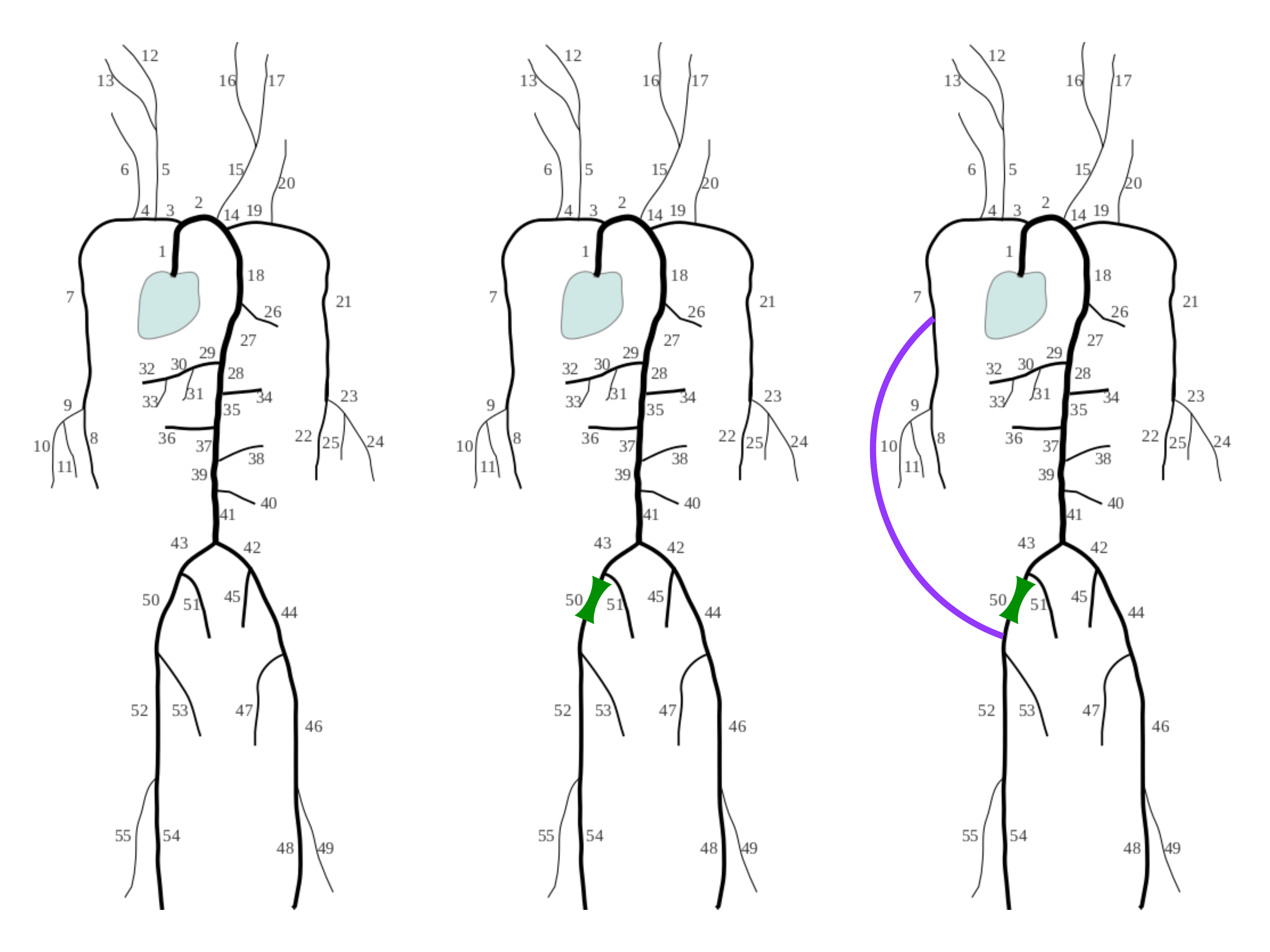}
	\caption{Arterial tree. (a) ``Healthy'' network. (b) Pathological network. (c) ``Repaired'' network. The pathological network (b) is modeled by  narrowing the cross-sectional area of the Right Iliac artery (number 50, green) and the extracorporeal bypass graft by an elastic tube (purple). In each segment, a 1D model of fluid flow with viscoelastic wall is solved numerically. The flow is imposed by given heart pulses, with a realistic reflection coefficient at the end of each terminal arteries. Table \ref{tab:systemic_network} presents the geometrical and mechanical data used in numerical computations.}
	\label{fig:Artree}
\end{figure}

The ``healthy'' network is based on the principal arteries of the great circulation (55 segments) and used in the literature as the basic model of the systemic network. Its topology is presented in Figure \ref{fig:Artree} (a), where each artery is given a number (ID) (useful to understand the numerical results), and the values of its arterial parameters are adapted from~\cite{Sherwin2003} and presented in Table~\ref{tab:systemic_network}. Compared to \cite{Sherwin2003}, we have added a viscoelastic term to the wall model which exists in reality and is very important from the hemodynamic point of view \cite{Alastruey2011,Wang2016}. In the literature on 1D network models, the viscoelastic term is usually not included as its coefficients are hard to evaluate experimentally. Here, we use the work of \cite{Armentano1995}, where the viscosity of the aortic walls of dogs was modeled by a Kelvin-Voigt model and where the values of $\phi$ range between $3.8 \pm 1.3 \times 10^3 \text{Pa}\cdot \text{s}$ and $7.8\pm1.1\times 10^3 \text{Pa}\cdot \text{s}$. Hence, in all numerical simulations we assume $\phi=5\times 10^3 \text{Pa}\cdot \text{s}$ to calculate the coefficient $C_v$. We note that in the absence of a viscoelastic term, high frequency components of the reflected signal would be present \cite{Wang2016}.

The hemodynamic flow through each segment is computed using the 1D numerical model presented in the previous section. The simulations are performed over ten heart periods and all data we show is taken in the final period. This is done to ensure that data is recorded only when a  permanent state is reached, where each heart period is identical to the next. The recorded data for the healthy  network contains the values of the blood flow rate $Q_{healthy}$, the arterial cross-sectional area $A_{healthy}$ and the blood pressure $P_{healthy}$ in every artery and at every recorded time of the final period. These numerical results are the target values used from now on to evaluate the pathological situation with respect to the healthy case and to assess the quality of the bypass grafts.

\subsection{Pathological case}

\subsubsection{Methods}

As stated in the introduction, the obliteration of an Iliac artery is a severe pathology since the Iliac arteries carry most of blood flow to the lower members. A reduction of the blood flow implies a low blood perfusion in distal sites of the lower leg and therefore a probable situation of ischemia. As explained before, we model the obliteration by narrowing the cross-sectional area of a portion of the Right Iliac artery (number 50 in Figure \ref{fig:Artree} (b)). The length of the occlusion is $5 \ cm$ and the degree of severity of the arterial occlusion is directly related to the ratio of the pathological value of the cross-sectional area of the stenosed artery over the value of the cross-sectional area of the healthy artery.

We define the ratio $I_s = { { A_{healthy} - A_{\%}} \over A_{healthy} } \times 100$ as the degree of obliteration. To assess the blood flow rate in the pathological network, we compare the computed blood flow rates to the healthy blood flow rate $Q_{healthy}$ obtained for $I_s=0 \%$. We note that in all presented results these values  are averaged in time over a complete heart cycle.  Four sites are chosen to evaluate the hemodynamical influence of the obliteration. Two are located in the lower legs, in the Right Femoral artery (number 52) and in its opposite  (Left Femoral, number 46). The other two are located in the arms, in the Right Subclavian artery (number 7)  and in its opposite  (Left Subclavian, number 21). The Right Femoral artery (number 52) is the principal assessment point of our numerical study because the flow rate passing through it determines the leg perfusion and therefore the degree of ischemia. The other arteries are used in clinical routines as control checks after a bypass graft surgery. 

\subsubsection{Results}

\begin{figure}[tbh]
\centering
\includegraphics[width=0.5\linewidth]{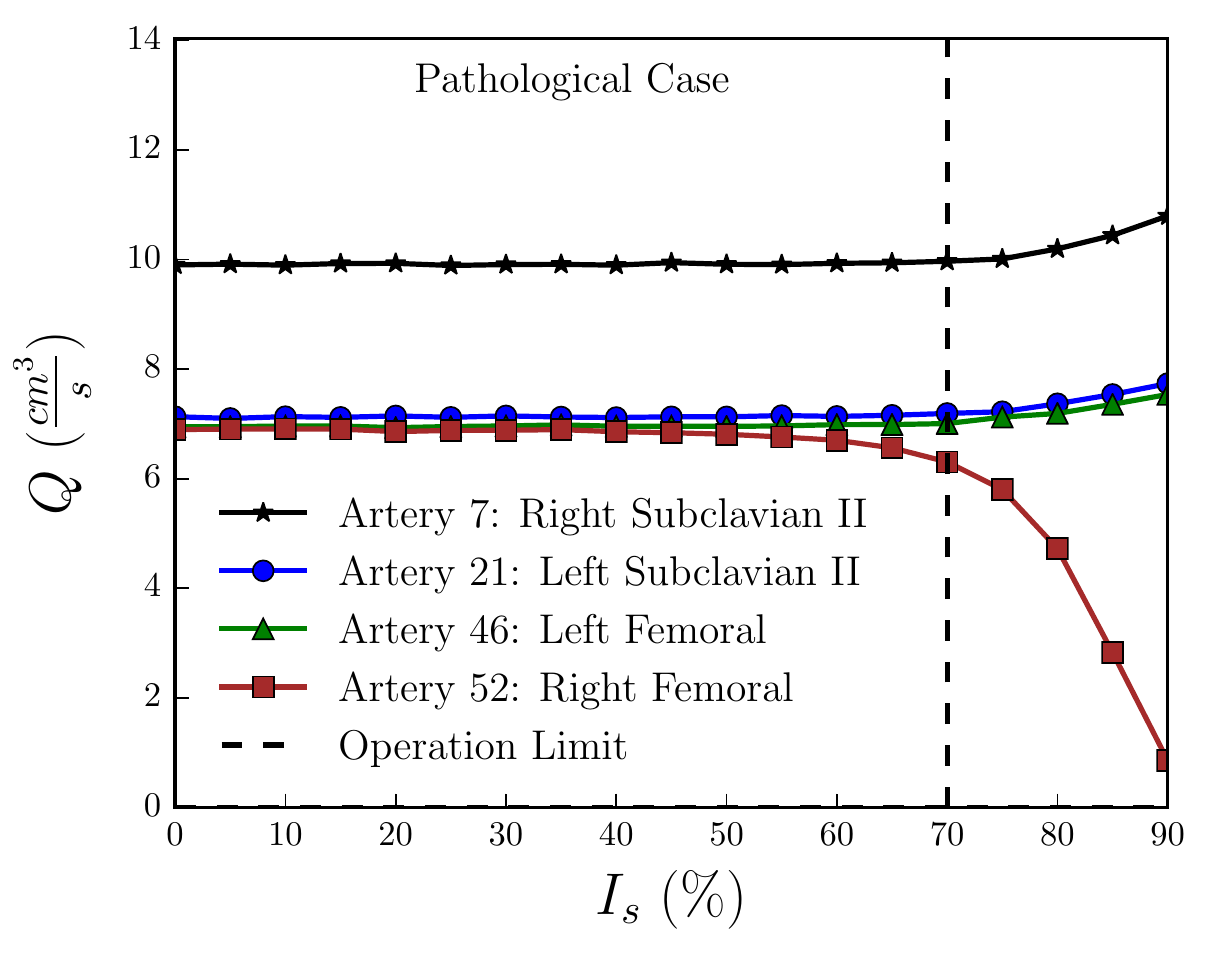}\includegraphics[width=0.5\linewidth]{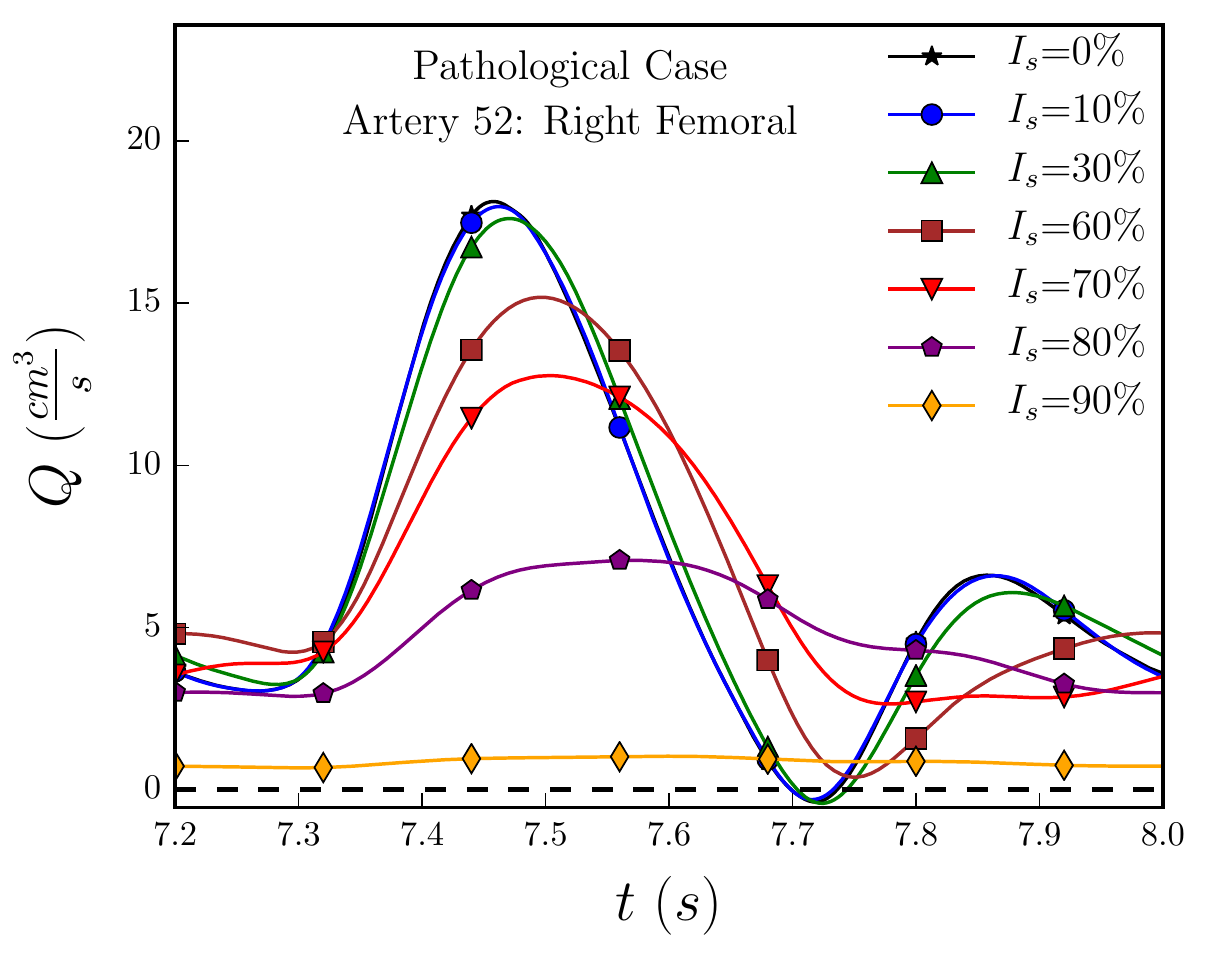}
\caption{(a) Flow rate averaged over a cycle as function of the obliteration degree $I_s$ for the following arteries: Right Femoral, (number 52), Left Femoral, (number 46), Right Subclavian (number 7)  and  Left Subclavian, (number 21). As the ratio $I_s$ increases, the flow rate drops in the Right Femoral artery, distal to the stenosis, whereas the flow rate increases in all other segments to compensate this drop. (b) Instantaneous flow rate as function of time over a cycle in the Right Femoral artery (number 52) for different degree of obliteration. As the ratio $I_s$ increases, the waveform looses its pulsatility and flattens and the average flow rate drops.}
\label{fig:fig3}
\end{figure}

 Figure \ref{fig:fig3} (a) shows the variation of the blood flow rate $Q$ with the index $I_s$ at the four previously defined control sites. The first observation is that we need between $60\% \ - \ 70 \%$ of narrowing to observe a significant variation of flow rate in the network, as is widely accepted by the medical community (i.e. renal arteries in pigs and human carotid arteries \cite{Lanzino2009,Rognant2010}). The Right Femoral artery (number 52) shows a drastic decrease in flow rate due to the obliteration of its proximal artery (number 51). We note that for an occlusion of $90 \%$ we have  almost no blood flow in the Right Femoral artery.  Conversely, the other measurement sites show a moderate rise of the blood flow rate to compensate the reduction of flow rate in the right leg distal to the stenosis. This as a clear example of how we can monitor global variations caused by a local mechanism. 

The Figure \ref{fig:fig3} (b) presents the time evolution of the blood flow rate $Q$ over a cycle after a transient state (after 9 heart cycles as stated before) in the Right Femoral artery (artery 52). The results are indeed correlated to those of the precedent figure (Figure \ref{fig:fig3} (a)) but notwithstanding gathering more significant information about the hemodynamics: first as expected the blood flow rate decreases in average as the ratio $I_s$ increases; second the positions of the maximum and minimum peaks are shifted, due to a time shift in the traveling waves; third the maximum amplitude decreases significantly as the ration $I_s$ increases and we observe that for $70 \%$ of occlusion the amplitude has dropped by $30 \%$ and for $80 \%$ of occlusion it drops by $60 \%$. This last point shows that as the degree of obliteration increases, the signal looses its pulsatility and flattens.
Indeed, we note that for $I_s = 90 \%$, the amount of blood perfusion in the leg is minimal (flat time line near zero in Figure \ref{fig:fig3} (b)).

\subsection{Bypasses}

\subsubsection{Methods}

A bypass graft consists basically of an elastic tube connecting two arteries, called donor and receptor. The donor artery is often a healthy artery and the choice of the receptor depends on the obstructed site.  The two connection points are called the proximal anastomosis, where one extremity of the bypass graft is attached to the donor artery, and the distal anastomosis, where the other extremity of the bypass graft is attached to the receptor artery. In our numerical model, we include the bypass graft by connecting an elastic tube between the donor and receptor arteries in the pathological network presented above. We compare blood flow rates and pressures over all the ``repaired'' network against the target data (the ``healthy'' network).  

\begin{figure}[!htb]
	\centering
	\includegraphics[width=0.7\textwidth]{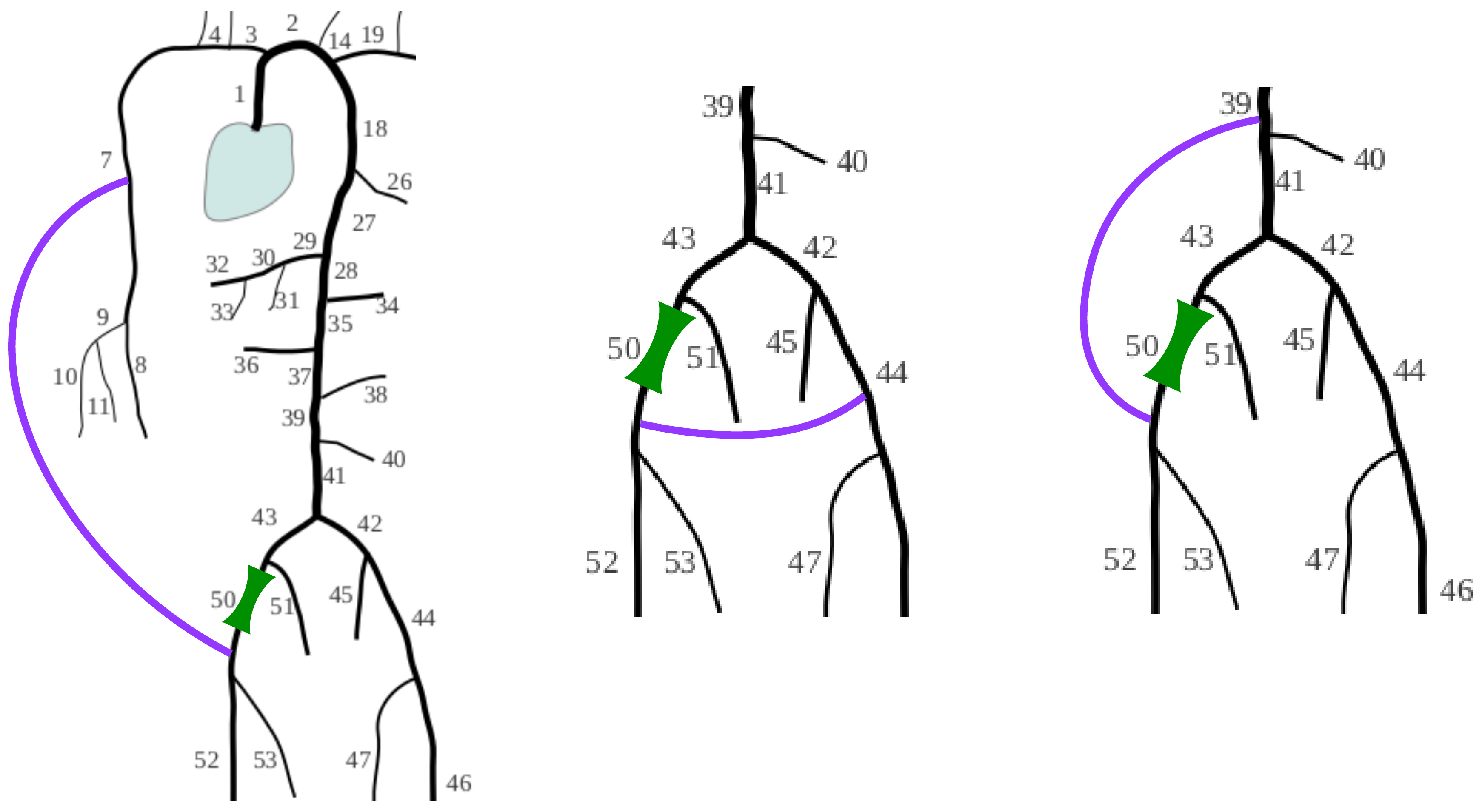}
	\caption{Sketch of three bypasses with the donor artery : (left) Axillo-Femoral (AxF) and donor artery,  Right Axillary artery (number 7), (center) cross-over-Femoral (FF) and donor artery, Left Femoral artery (number 44) and (right) Aorto-Femoral (ArF) and donor artery, Abdominal Aorta (numer 39).}
	\label{fig:bypasses}
\end{figure}

We study three different bypass grafts, named with respect to their donor and receptor sites, the receptor being in all cases the distal part of the Right Iliac artery (artery 51) just after the occlusion site. In Figure \ref{fig:bypasses} we have sketched the topology of the three different bypass grafts: Axilo-Femoral (AxF) where the donor artery is the Axillary artery (artery 7), cross-over Femoral (FF) where the donor artery is the opposite Common Femoral artery (artery 44) and last Aorto-Femoral (ArF) where the donor artery is the Abdominal Aorta (artery 39).


The mechanical and geometrical characteristics of each bypass graft, such as the compliance, length and diameter, are taken from the literature \cite{Sarkar2006}. Each bypass graft we study is made of the same composite material which is constituted principally of polyethylene terephthalate (Dacron), with a Young's modulus equal to $9 \ 10^{6} Pa$, an internal diameter of  $0.8 cm$ and a thickness of $0.05 \ cm$. The bypass grafts lengths $L$ follow the geometric distances between the donor and receptor sites:  Axillo-femoral, $40 \ cm$, cross-over-femoral, $20 \ cm$  and Aorto-femoral, $20 \ cm$.

It is important to notice that the numerical simulations give access to all dynamic variables (pressure, cross-sectional area, flow rate) simultaneously over the entire network, and that these specific data are impossible to obtain in clinical routines. We recall the main objective of this numerical study, (i) to assess the new hemodynamic conditions for each bypass graft to understand the modified flow hemodynamics and then (ii) to help medical staff decide which bypass graft is better. 

To assess the performances of each bypass graft, we define three control sites in which we will compare the numerical data of the ``healthy'', pathological and ``repaired'' cases. The first one is the Right Femoral artery, downstream of the obliteration and the distal anastomosis, and is identical to the control site used to analyze the pathological network. The second and third control sites are respectively the upstream and downstream segments of the proximal anastomosis, that is the point where the bypass graft is connected to the donor artery.

\subsubsection{Results}
  
\begin{figure}[!htb]
	\centering
	\includegraphics[width=0.5\linewidth]{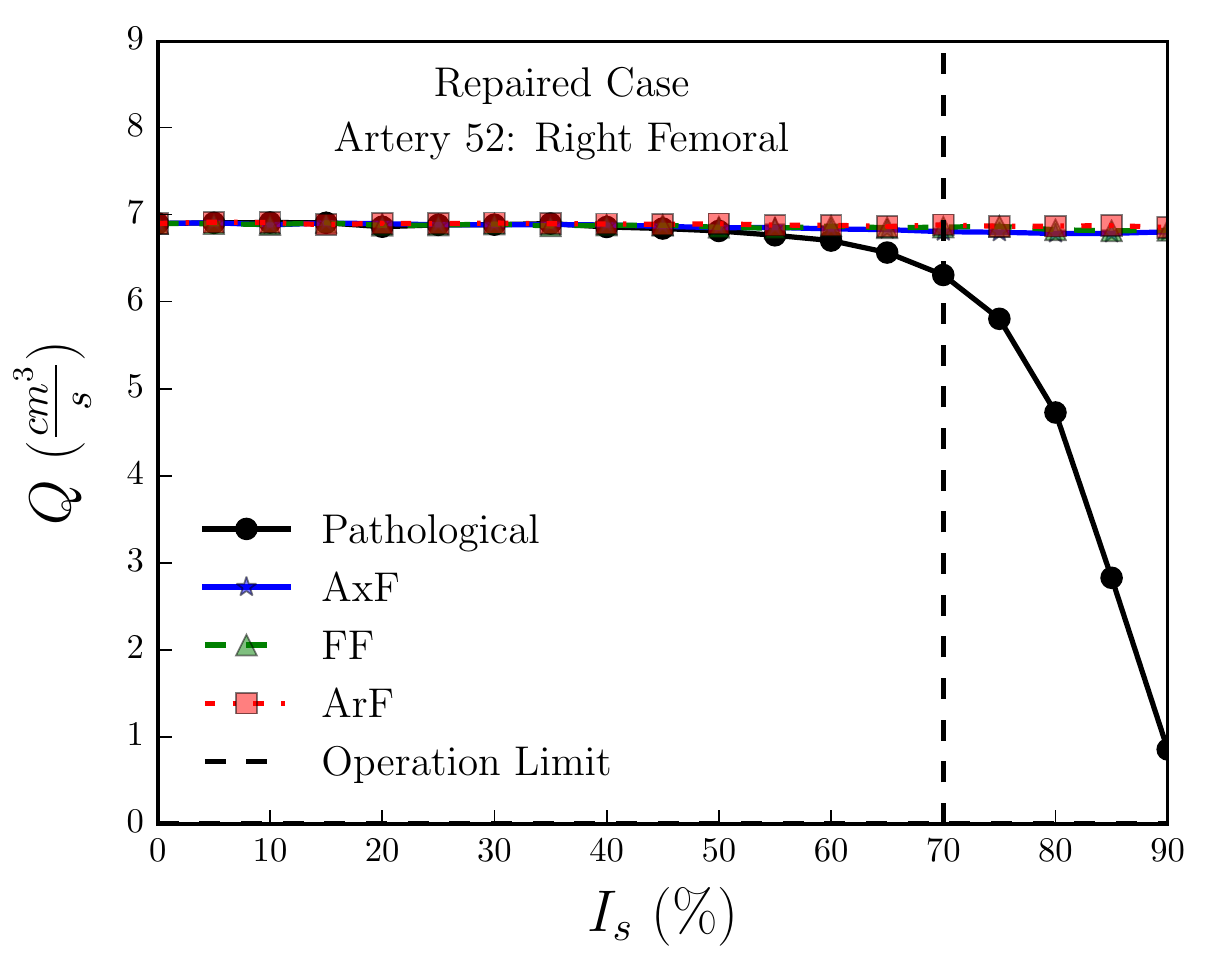}\includegraphics[width=0.5\linewidth]{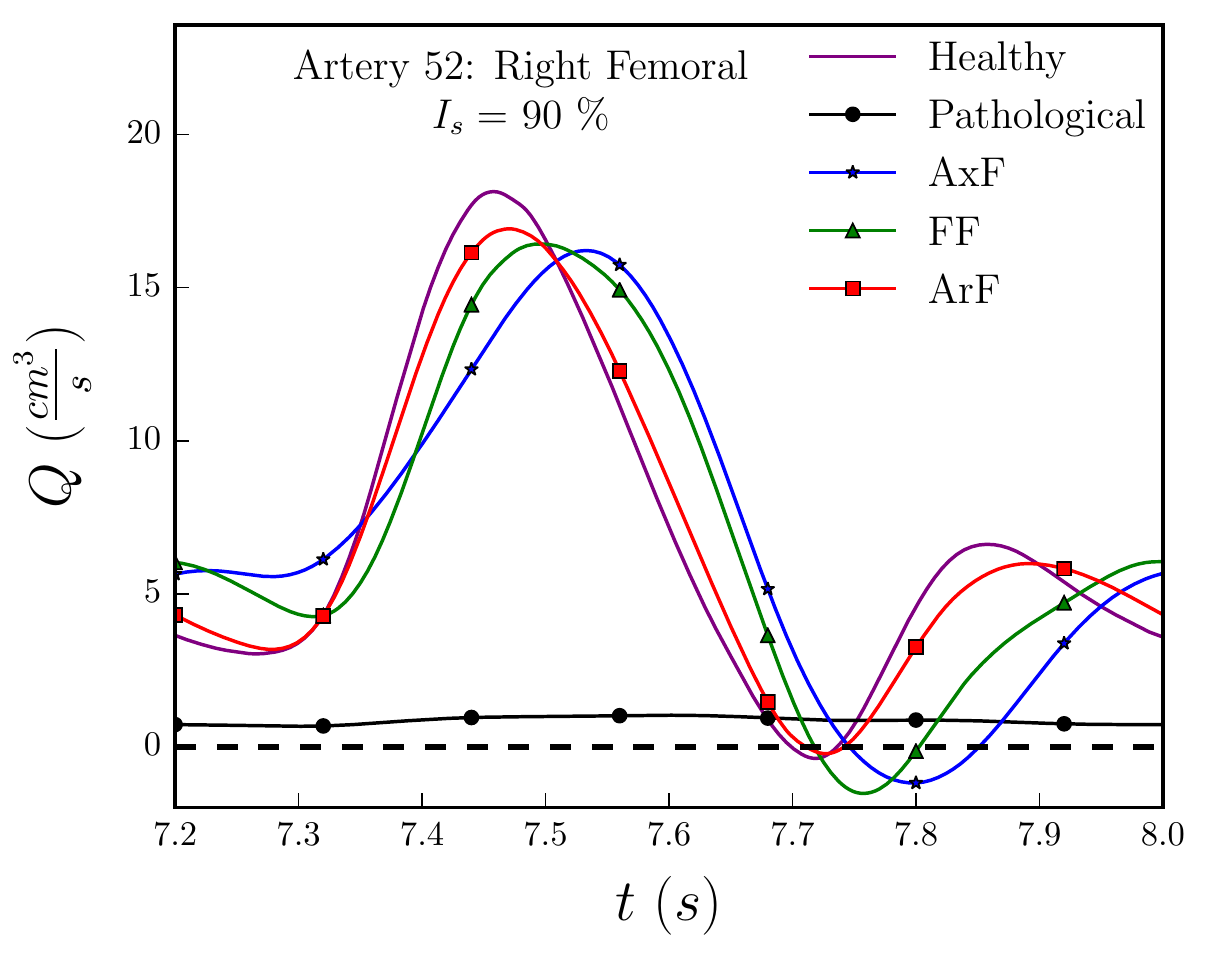}
	\caption{(a) Averaged flow rate over a cycle as function of the degree of obliteration $I_s$ (Artery 52 : Right Femoral) (b) Instantaneous flow rate as function of time over a cycle (Artery 52 : Right Femoral) for healthy, pathological with  $I_s=90 \%$ and for the three bypasses. We observe that for all three bypass graft configurations, we are able to recover the target healthy flow rate (average values and waveform) distal to the stenosis.}
	\label{fig:Multiplot_all}
\end{figure}

We first study for the three bypass grafts the predicted perfusion hemodynamics in the first control site located after the stenosis, in the Right Femoral artery (number 52). The Figure \ref{fig:Multiplot_all} (a) presents the numerical results for the time-averaged blood flow rate for the pathological state (same as Figure \ref{fig:fig3} (a) for the artery 52) and for the repaired states using the three different bypass grafts. Figure \ref{fig:Multiplot_all} (b) shows the temporal evolution of the blood flow rate over one heart cycle after the transitory state for $I_s = 90 \ \%$. These figures should be compared to Figures \ref{fig:fig3} (a) and (b).  We observe in Figure \ref{fig:Multiplot_all} (a) that for all three bypass grafts configurations, we retrieve in average the blood flow rate of the healthy case for every degree of obliteration considered. Figure \ref{fig:Multiplot_all} (b) shows that the waveform of the blood flow rate is similar to the target wave form (Healthy) although the amplitude of the peaks are a bit mis-estimated.  Furthermore there is a time-shift in the position of the maximum and minimum flow rate peaks due to the fact that the length of vessels traveled by the wave differs from the healthy case in each bypass graft configuration. Even if the blood flow rate peaks are not exactly the same,  we retrieve for all three bypass grafts configurations the target average blood flow rate as well as the approximate shape of the waveform. From the analysis of Figure \ref{fig:Multiplot_all}  we can conclude that all three bypass graft configurations are successful in retrieving the healthy flow rate in the first control site distal to the obliterated segment (the Right Femoral artery 52).

We complete our study of each bypass graft configuration by analyzing the blood flow rate in each donor artery. In the subsequent numerical results we analyze in  particular two control sites of the donor artery: the upstream and downstream segments of the proximal anastomosis placed before and after the bypass connection point respectively. We expect that the bypass will (i) supply the missing blood flow rate to the unwell lower leg (Right Femoral artery 52) and (ii) be able to continue a healthy perfusion at the donor site (downstream segments of the proximal anastomosis).

For the cross-over Femoral bypass graft (FF) the  donor artery is the opposite Femoral artery (Left Femoral artery, number 44 in Figure \ref{fig:Artree}). 

\begin{figure}[!htb]
\centering
\includegraphics[width=0.5\linewidth]{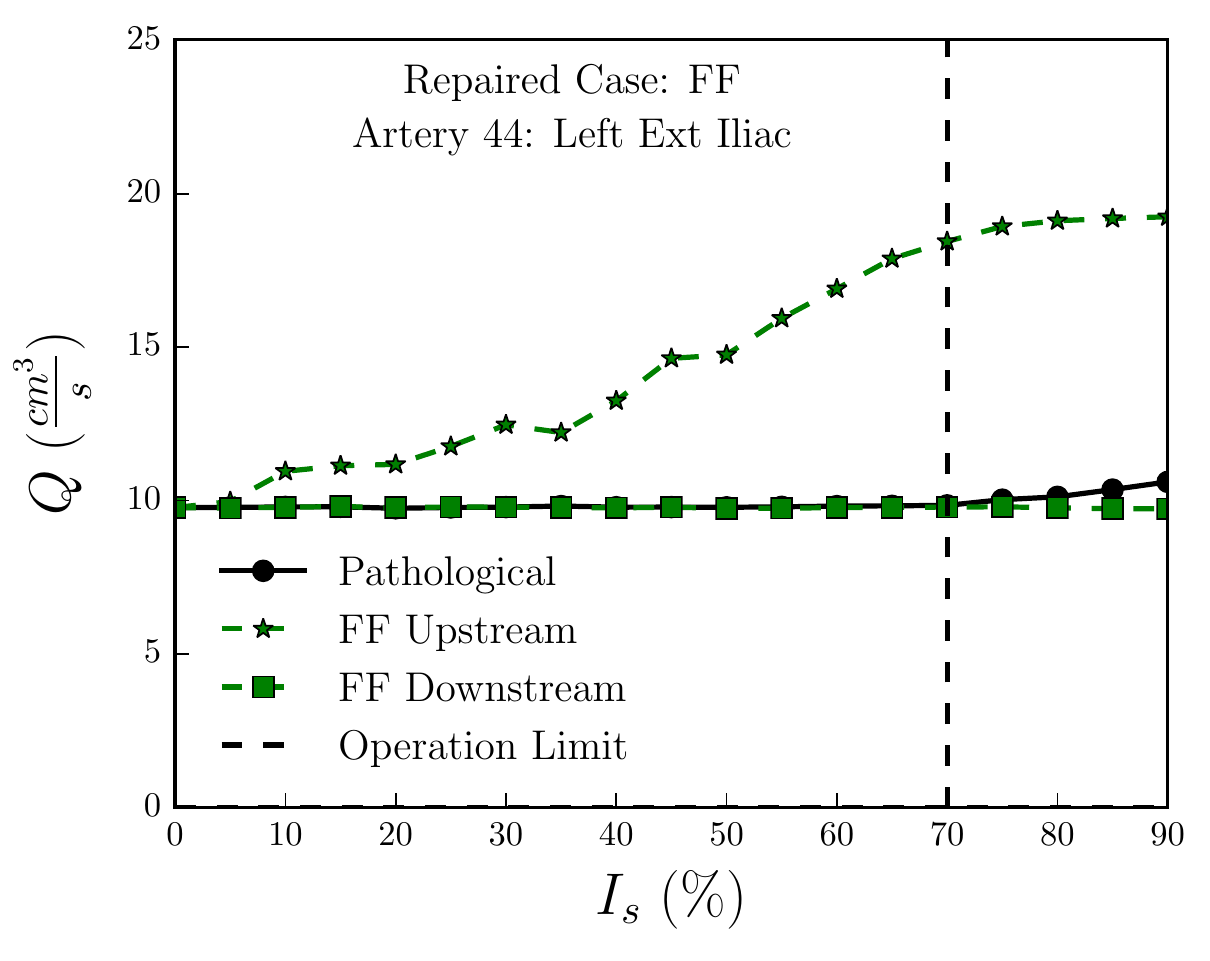}
\caption{Cross-over Femoral bypass graft: average flow rate over a cycle in the opposite Femoral artery (Artery 44). Upstream of the proximal anastomosis, the flow rate increases to properly vascularize the bypass graft, depending on the degree of obliteration $I_s$. Downstream of the proximal anastomosis, we recover the healthy ($I_s = 0 \%$) flow rate.}
\label{fig:Bypass44}
\end{figure}

The Figure \ref{fig:Bypass44} presents the computed blood flow rate as function of the degree of obliteration $I_s$ in the donor artery (Left Femoral artery) at the two control sites, upstream and downstream of the proximal anastomosis. We observe that upstream of the donor site in the Left Femoral artery (artery 44) the flow rate rises proportionally to the degree of obliteration. This is the signature that the donor artery must know supply blood to both the downstream segment as well as the unwell member and therefore increase its flow rate, in comparison to the healthy case ($I_s = 0 \%$). This behavior shows that the cross-over Femoral bypass graft is carrying correctly the blood to the stenosed member. The downstream blood flow rate does not change compared to the healthy case ($I_s = 0 \%$) indicating that the opposite lower leg, downstream of the proximal anastomosis, is correctly supplied. We note that for a severe stenosis (obliteration of 90 \%) the upstream blood flow rate is twice the basal one.

\begin{figure}[!htb]
\centering
\includegraphics[width=0.5\linewidth]{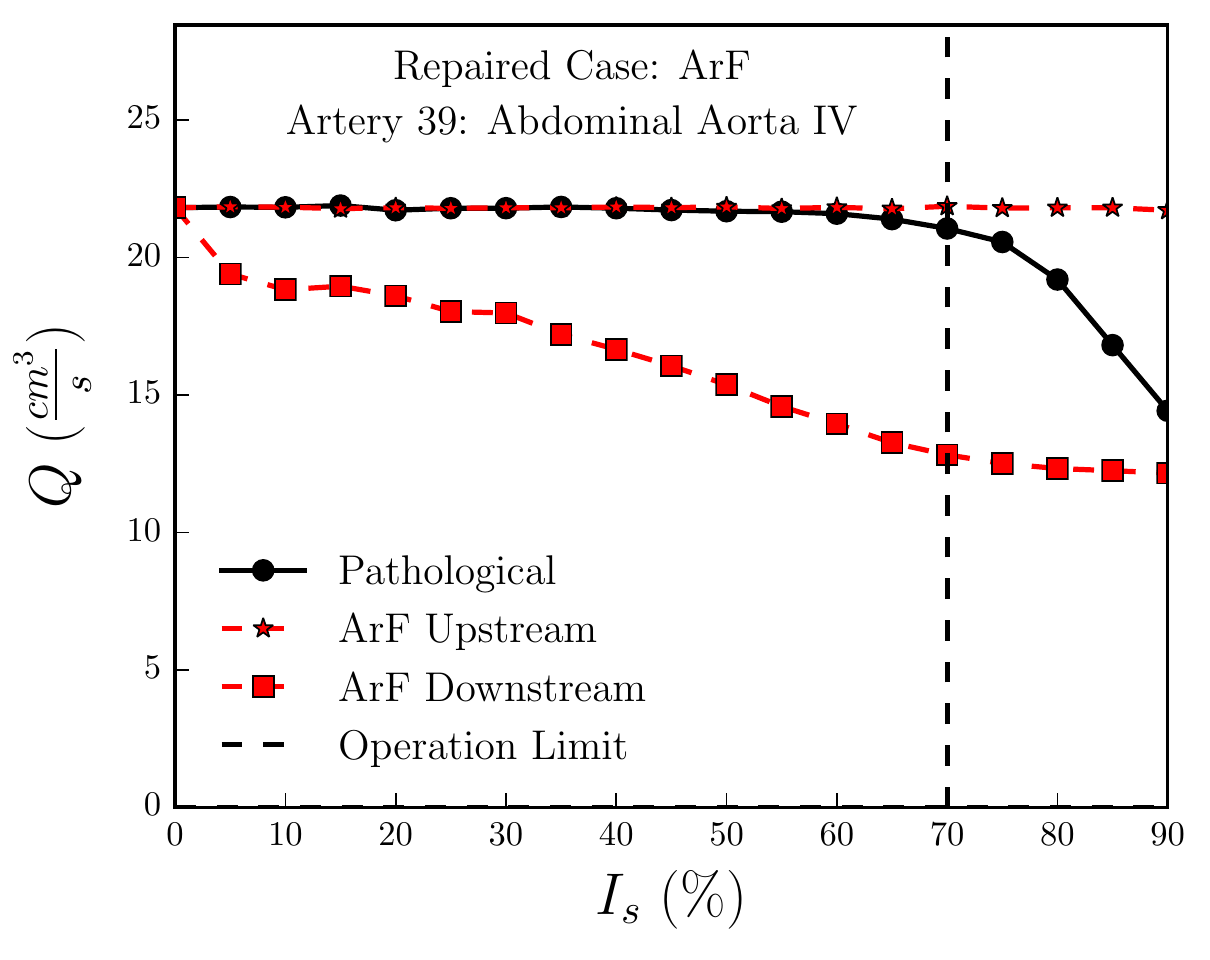}
\caption{Aorto-Femoral bypass graft: average flow rate over a cycle at the donor  artery (Artery 39). Upstream of the proximal anastomosis, the flow rate remains unchanged since blood flow passing through the bypass graft to vascularize the right leg was already supplied by the Aorta in the healthy case. Downstream of the proximal anastomosis, the flow rate decreases with the degree of obliteration $I_s$ since now only the blood supplying the left leg is passing downstream of the proximal anastomosis.}
\label{fig:Bypass38}
\end{figure}

For the Aorto-Femoral bypass graft (ArF) the donor artery is the Abdominal Aorta (artery 39), the principal path carrying the blood to both lower legs. We observe in Figure \ref{fig:Bypass38} that upstream of the donor site in the Abdominal Aorta (artery 39), the blood flow rate does not change with the degree of obliteration, contrary to the previous cross-over Femoral (FF) bypass graft configuration. Indeed, in the healthy configuration the Abdominal Aorta already caries blood the the Right Femoral artery, contrary the donor artery of the previous bypass graft configuration, the Left Femoral artery (artery 44). Therefore no compensation mechanism is required in presence of the bypass upstream of the donor site. Conversely we observe that the downstream of the proximal anastomosis, the blood flow rate decreases as the degree of obliteration increases, in comparison to the healthy configuration ($I_s = 0 \%$). Indeed, since the blood that vascularizes the unwell member (Right Femoral artery 52) now flows through the bypass graft, only the blood supply of the left leg remains downstream of the donor site. This behavior shows that the bypass graft is indeed bringing blood the the stenosed member. To conclude the analysis, we can note that in absence of stenosis ($0 \%$) the downstream blood flow is symmetrically shared between the two legs and that for a severe stenosis (obliteration of $90 \%$) the downstream blood flow rate is half the basal one.

\begin{figure}[!htb]
\centering
\includegraphics[width=0.5\linewidth]{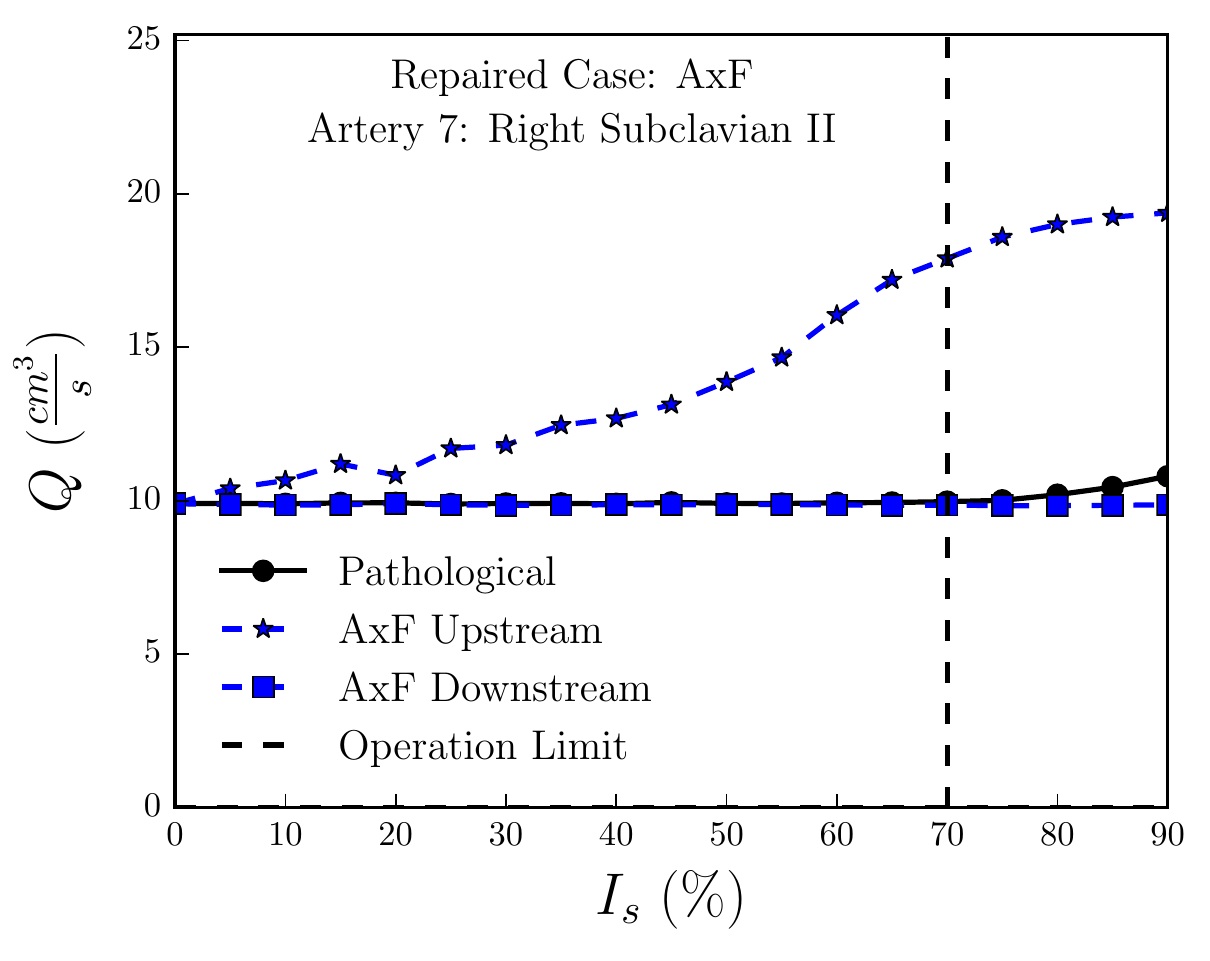}
\caption{Axillo-Femoral bypass graft: averaged flow rate over a cycle at the donor artery (number 7). Upstream of the proximal anastomosis, the flow rate increases to properly vascularize the bypass graft, depending on the degree of obliteration $I_s$. Downstream of the proximal anastomosis, we recover the healthy ($I_s = 0 \%$) flow rate.}
\label{fig:Bypass7}
\end{figure}

The Figure \ref{fig:Bypass7} presents the evolution of the blood flow rate upstream and downstream of the donor site as a function of the degree of obliteration varies for the last bypass graft configuration, the Axillo-Femoral bypass graft (AxF). The behavior of this graft is identical to the one of the cross-over Femoral (FF) bypass graft. The same analysis can be performed and we can conclude that this bypass graft configuration allows to correctly supply the unwell member while maintaining the healthy flow rate downstream of the donor site. We also note hat for an obliteration of $90 \%$ the upstream blood flow rate is twice the basal one.

In all three bypass graft configurations, the target behaviors were obtained and the bypass graft surgery was successful.

\subsection{Optimization}

The Axillo-Femoral bypass graft (AxF) which connects the Axillary artery (donor artery) to the obliterated artery (Figure \ref{fig:Artree} (c) and Figure \ref{fig:bypasses} (a)) is different from the other two bypass grafts we consider. The elastic tube used to connect the donor and receptor arteries is placed under the skin. Indeed, patients who undergo this type of surgery are not healthy enough to survive the more invasive surgical procedures required by the other two bypass grafts. Moreover, its donor and receptor arteries are the furthest away. It is therefore the longest bypass graft, which increases the chances of graft failure.

\begin{figure}[!htb]
\centering
\includegraphics[width=0.7\linewidth]{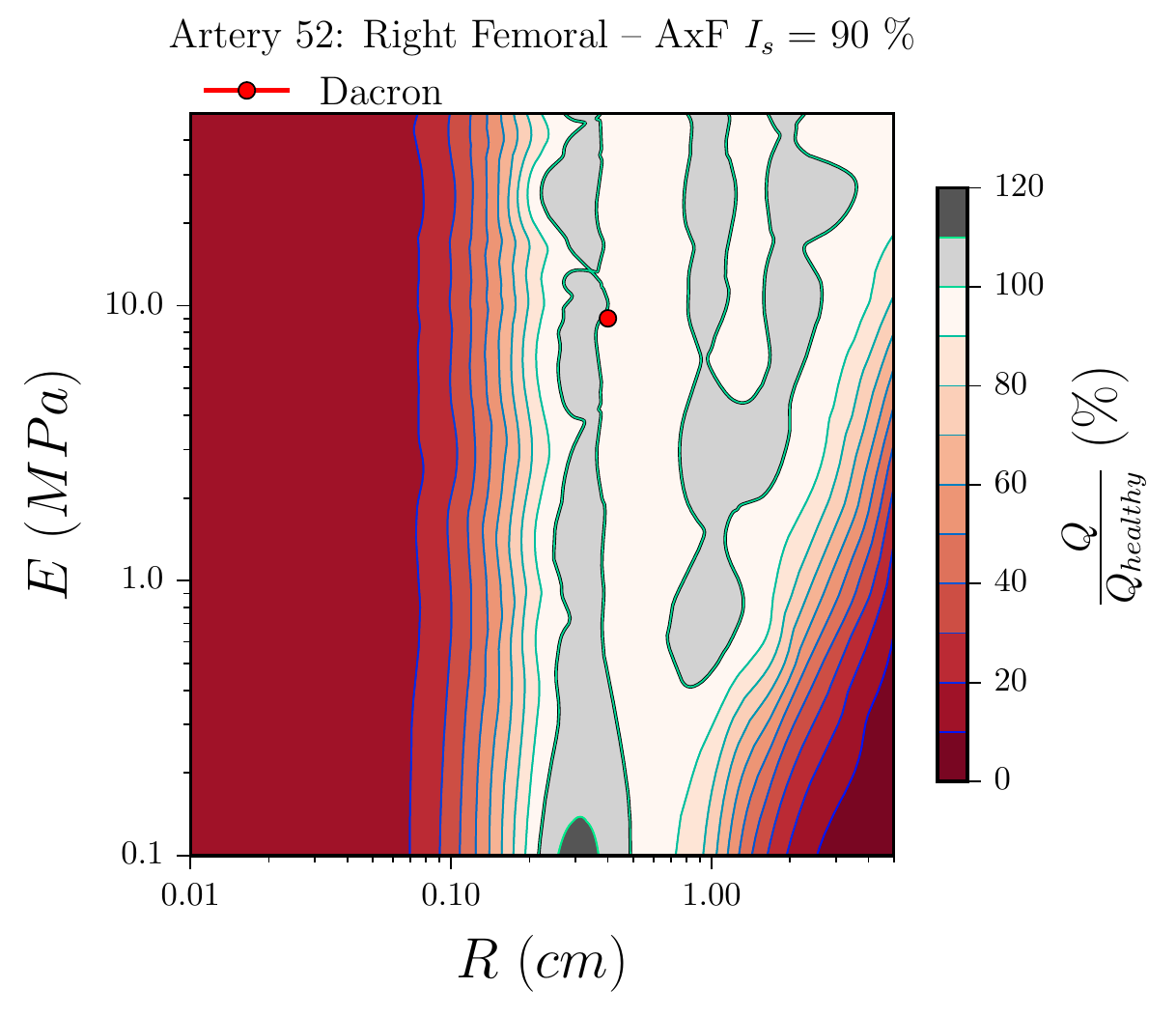}
\caption{Contour plot for the normalized averaged flow rate ${ Q \over Q_{healthy}}$  as function of the Young modulus $E$ and the radius $R$ for a stenosis of $90 \ \%$. The red circle corresponds to the actual values of the Young modulus $E$ and the radius $R$ used in numerical simulations, which are situated in an optimal zone ($100 \%$). When the radius decreases, the resistance of the tube increases and therefore less flow is bypassing through the bypass graft. When the Young's modulus decreases, the tube becomes more compliant and stores more flow. Both behaviors reduce the quality of the bypass graft.}
\label{fig:Contour_ER_Q}
\end{figure}

\begin{figure}[!htb]
\centering
\includegraphics[width=0.7\linewidth]{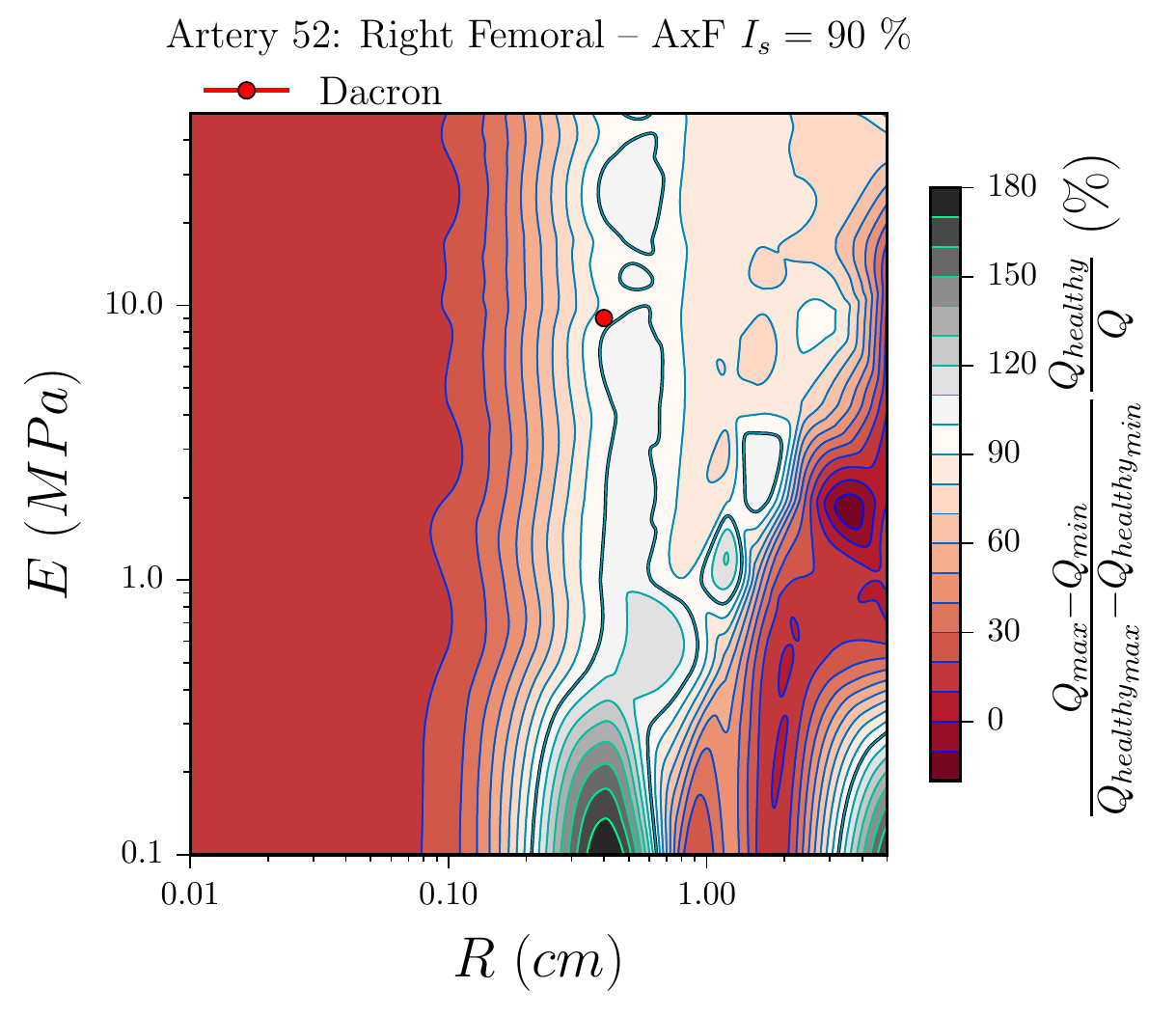}
\caption{Contour plot for the normalized peak to peak flow rate ${ Q_{max} - Q_{min} \over Q^{healthy}_{max} - Q^{healthy}_{min}} {Q_{healthy} \over Q}$ as function of the Young modulus $E$ and the radius $R$ for a stenosis of $90 \ \%$. The red circle corresponds to the actual values of the Young modulus $E$ and the radius $R$ used in numerical simulations.}
\label{fig:Contour_ER_max-min}
\end{figure}

For these reasons, we choose to perform a more detailed analyze of the Axillo-Femoral bypass graft (AxF) by asking the question: are the mechanical (the Young modulus $E$) and geometrical (the radius $R$) parameters optimal in some sense?
In order to give arguments for discussion we evaluate the quality of the mechanical characteristics of the bypass graft by performing hundreds of simulations in which we vary the values of the Young modulus [$0.1 - 50 \ MPa$] and the radius $R$ [$0.01 - 5 \ cm$]. As before, we used as a target the healthy data in the Right Femoral artery (numbered 51) to asses re-perfusion due to the presence of the bypass. Figures \ref{fig:Contour_ER_Q} and \ref{fig:Contour_ER_max-min} present the normalized flow rate ${ Q \over Q_{healthy}}$ and the normalized peak to peak flow rate ${ Q_{max} - Q_{min} \over Q^{healthy}_{max} - Q^{healthy}_{min}} {Q_{healthy} \over Q}$ respectively for an obliteration degree of $90 \ \%$. Data for the ``healthy'' network give the reference values.

The circle in the middle of the Figure \ref{fig:Contour_ER_Q} indicates the actual values of the bypass graft's Young's modulus and radius (Note that the axis are in a log-log scale). This point is inside of the zone of 100\% in terms of the normalized flow rate ${ Q \over Q_{healthy}}$ showing that these values allow to restore in average the healthy flow rate in the lower leg. Starting from the bypass graft point we analyze the results by moving along the horizontal and vertical directions, that is for $E$ constant and varying $R$ (horizontal) and the opposite, for $R$ constant and varying $E$ (vertical). Moving in the horizontal direction towards the left we are reducing the radius $R$ of the bypass,  hence increasing the hydraulic resistance. Consequently the flow rate goes quickly down. If we move towards the right the numerical results seem less clear, as we are increasing the radius so increasing the blood volume inside of the bypass. Even if the distal blood perfusion is still correct it is clear that the proximal site of the bypass will receive much less flow rate and there is a possible risk of ischemia in the right hand. Another mechanical argument is that increasing the radius implies decreasing blood flow velocity which will result in a smaller shear rate along the bypass. A smaller shear rate inside of the bypass graft is a major drawback since it increases aggregation and coagulation processes.  From a physiological and mechanical point of view for a given Young modulus $E$, the optimal radius of the bypass graft should be as close as possible to a 100\% zone and as small as possible (towards the left), ensuring in this way an accurate distal and proximal blood perfusion.

For a constant radius $R$ we analyze the Young modulus variations. When considering wave reflections, the bypass graft's Young's modulus should be close to the arteries' Young's modulus since elasticity jumps lead to impedance discontinuities and therefore higher reflected pressure waves. If the bypass elasticity is lower than the arterial one (moving downwards from the circle), the bypass graft will become more compliant and inflate, increasing again the blood volume inside inside the bypass graft. Conversely for stiffer bypass grafts (moving upwards from the circle) the distal flow rate in artery 51 will be reliable but we will generate high pressure peaks du to the increased reflections. 

Figure \ref{fig:Contour_ER_max-min} presents the  normalized peak to peak flow rate ${ Q_{max} - Q_{min} \over Q^{healthy}_{max} - Q^{healthy}_{min}} {Q_{healthy} \over Q}$. This quantity measures the pulsatility of the flow rate signal (and consequently the pressure signal). It helps to better understand the above discussion about the normalized flow rate. We observe that both figures are quite similar, and the previous description for a fixed Young's modulus $E$ and radius $R$ can be applied. Nevertheless the high radius zone for fixed Young's modulus is a bit different. We previously described this situation as a correlation between an increase of radius and an increase of the blood volume inside the bypass graft. In such a zone, Figure \ref{fig:Contour_ER_max-min} shows that the quality of the bypass graft falls because the signal amplitude becomes smaller and smaller, as blood is being stored inside the bypass graft.

\section{Conclusion}

We have presented a 55 arteries network with viscoelactic arteries deformed by the blood flow  modeled by 1D fluid-structure system of equations. We have done simulations of this complex nonlinear dissipative system in a network with a Right Iliac artery obliteration. 
We computed the blood flow in the arterial network where the three classical bypass grafts for an iliac obliteration were used, using the network parameters of an ``average'' man and the mechanical characteristics of the bypass graft were taken from the literature.

Numerical simulations show that all bypass grafts have similar success since we retrieve the healthy hemodynamics after bypass graft surgery. Moreover the control checkout on the donor arteries of each bypass was satisfied. 
 
 We have studied in particular the optimization of the Axillo-Femoral bypass because it is proposed for unhealthy weak patients who can not sustain other types of bypass graft surgeries and because it has the less graft survival time of the three bypasses studied. Little is known about the hemodynamic evolution as a function of mechanical characteristic of the bypass grafts, and the optimization results (Figures \ref{fig:Contour_ER_Q} and \ref{fig:Contour_ER_max-min}) have shown that the classical mechanical characteristic of the grafts used for evaluation are optimal in terms of the hemodynamic quantities. This result suggests that both the mechanical characteristic and the diameter of the bypass graft are well dimensioned, because other set of parameters will lead to non-restored hemodynamics. We note also that the explored parameter domain is quite consequent (2,5 orders of magnitude in Young's modulus and radius) which enhances the pertinence of the numerical study. 

Besides the numerical approach, the numerical findings over an "averaged patient" prove that numerical hemodynamic predictions could be used to optimize or plan surgeries for specific patients, under the conditions that the pathologies were well defined and the physiological parameters known. Another important point is that the numerical tool is very fast in terms of computing time,  and could be used to do massive simulations (for instance doing parameter's statistics and using technique of error propagation) and to numerically evaluate the performances of new grafts. 

\pagebreak
 
\appendix

\section{Wall model and geometrical and mechanical parameters of the network}
\label{annexe}

The Table \ref{tab:systemic_network} presents the name (ID), name, length, neutral cross-sectional area $A_0$ and mechanical parameters used in the network. 

\begin{longtable}{c c c c c c c} 
	\caption{Arterial network: Data adapted from~\cite{Sherwin2003} and \cite{Armentano1995}}
	\tabularnewline \hline 
		   &      &  $l$          & $A_0$           & $\beta$              & $C_v$  & \tabularnewline
		ID & Name & $(\text{cm})$ & $(\text{cm}^2)$ & $(10^6 \text{Pa/cm})$& $(10^4 \text{cm}^2\text{/s})$ & $R_t$ \tabularnewline \hline \endhead
	\hline
	 \endfoot
	\hline
	\hline
	1 & Ascending aorta & 4.0 & 6.789 & 0.023 & 0.352 & --\\
	2 & Aortic arch I & 2.0 & 5.011 & 0.024   & 0.317 & -- \\
	3 & Brachiocephalic & 3.4 & 1.535 & 0.049 & 0.363 & -- \\
	4 & R.subclavian I & 3.4 & 0.919 & 0.069  & 0.393 & -- \\
	5 & R.carotid & 17.7 & 0.703 & 0.085      & 0.423 & -- \\
	6 & R.vertebral & 14.8 & 0.181 & 0.470    & 0.595 & 0.906 \\
	7 & R. subclavian II & 42.2 & 0.833 & 0.076 & 0.413 & -- \\
	8 & R.radius & 23.5 & 0.423 & 0.192       & 0.372 & 0.82 \\
	9 & R.ulnar I & 6.7 & 0.648 & 0.134       & 0.322 & -- \\
	10 & R.interosseous & 7.9 & 0.118 & 0.895 & 0.458 & 0.956 \\
	11 & R.ulnar II & 17.1 & 0.589 & 0.148    & 0.337 & 0.893\\
	12 & R.int.carotid & 17.6 & 0.458 & 0.186 & 0.374 & 0.784 \\
	13 & R. ext. carotid & 17.7 & 0.458 & 0.173 & 0.349 & 0.79 \\
	14 & Aortic arch II & 3.9 & 4.486 & 0.024 & 0.306 & -- \\
	15 & L. carotid & 20.8 & 0.536 & 0.111    & 0.484 & -- \\ 
	16 & L. int. carotid & 17.6 & 0.350 & 0.243 & 0.428 & 0.784 \\
	17 & L. ext. carotid & 17.7 & 0.350 & 0.227 & 0.399 & 0.791 \\
	18 & Thoracic aorta I & 5.2 & 3.941 & 0.026 & 0.312 & --\\
	19 & L. subclavian I & 3.4 & 0.706 & 0.088  & 0.442 & -- \\
	20 & L. vertebral & 14.8 & 0.129 & 0.657    & 0.704 & 0.906 \\
	21 & L. subclavian II & 42.2 & 0.650 & 0.097 & 0.467 & -- \\
	22 & L. radius & 23.5 & 0.330 & 0.247       & 0.421 & 0.821 \\
	23 & L. ulnar I & 6.7 & 0.505 & 0.172       & 0.364 & -- \\
	24 & L. interosseous & 7.9 & 0.093 & 1.139  & 0.517 & 0.956 \\
	25 & L. ulnar II & 17.1 & 0.461 & 0.189     & 0.381 & 0.893 \\
	26 & intercoastals & 8.0 & 0.316 & 0.147    & 0.491 & 0.627 \\
	27 & Thoracic aorta II & 10.4 & 3.604 & 0.026 & 0.296 & -- \\
	28 & Abdominal aorta I & 5.3 & 2.659 & 0.032 & 0.311 & -- \\
	29 & Celiac I & 2.0 & 1.086 & 0.056        & 0.346 & -- \\
	30 & Celiac II & 1.0 & 0.126 & 0.481       & 1.016 & -- \\
	31 & Hepatic & 6.6 & 0.659 & 0.070         & 0.340 & 0.925 \\
	32 & Gastric & 7.1 & 0.442 & 0.096         & 0.381 & 0.921 \\
	33 & Splenic & 6.3 & 0.468 & 0.109         & 0.444 & 0.93 \\
	34 & Sup. mensenteric & 5.9 & 0.782 & 0.083 & 0.439 & 0.934 \\
	35 & Abdominal aorta II & 1.0 & 2.233 & 0.034 & 0.301 & -- \\
	36 & L. renal & 3.2 & 0.385 & 0.130        & 0.481 & 0.861 \\ 
	37 & Abdominal aorta III & 1.0 & 1.981 & 0.038 & 0.320 & -- \\
	38 & R. renal & 3.2 & 0.385 & 0.130        & 0.481 & 0.861 \\ 
	39 & Abdominal aorta IV & 10.6 & 1.389 & 0.051 & 0.358 & -- \\ 
	40 & Inf. mesenteric & 5.0 & 0.118 & 0.344  & 0.704 & 0.918 \\
	41 & Abdominal aorta V & 1.0 & 1.251 & 0.049  & 0.327 & -- \\
	42 & R. com. iliac & 5.9 & 0.694 & 0.082   & 0.405 & -- \\
	43 & L. com. iliac & 5.8 & 0.694 & 0.082   & 0.405 & -- \\
	44 & L. ext. iliac & 14.4 & 0.730 & 0.137  & 0.349 & -- \\
	45 & L. int. iliac & 5.0 & 0.285 & 0.531   & 0.422 & 0.925 \\
	46 & L. femoral & 44.3 & 0.409 & 0.231     & 0.440 & -- \\
	47 & L. deep femoral & 12.6 & 0.398 & 0.223 & 0.419 & 0.885 \\
	48 & L. post. tibial & 32.1 & 0.444 & 0.383 & 0.380 & 0.724 \\
	49 & L. ant. tibial & 34.3 & 0.123 & 1.197  & 0.625 & 0.716 \\
	50 & L. ext. iliac & 14.5 & 0.730 & 0.137   & 0.349 & -- \\ 
	51 & R. int. iliac & 5.0 & 0.285 & 0.531    & 0.422 & 0.925 \\
	52 & R. femoral & 44.4 & 0.409 & 0.231      & 0.440 & -- \\ 
	53 & R. deep femoral & 12.7 & 0.398 & 0.223 & 0.419 & 0.888 \\
	54 & R. post. tibial & 32.2 & 0.442 & 0.385 & 0.381 & 0.724 \\
	55 & R. ant. tibial & 34.4 & 0.122 & 1.210  & 0.628 & 0.716 \\
	\hline
	\label{tab:systemic_network}
\end{longtable}


\pagebreak
\fancyfoot{}
\fancyhead{}
\renewcommand{\headrulewidth}{0pt}
\renewcommand{\footrulewidth}{0pt}

\bibliography{\myreferences}


\end{document}